\documentclass[default,iicol]{sn-jnl}% Default with double column layout

\usepackage{graphicx}%
\usepackage{multirow}%
\usepackage{amsmath,amssymb,amsfonts}%
\usepackage{amsthm}%
\usepackage{mathrsfs}%
\usepackage[title]{appendix}%
\usepackage{xcolor}%
\usepackage{textcomp}%
\usepackage{manyfoot}%
\usepackage{booktabs}%
\usepackage{algorithm}%
\usepackage{algorithmicx}%
\usepackage{algpseudocode}%
\usepackage{listings}%
\usepackage{subfig}

\newcommand{\git}{\mathbin{
  \mathchoice{/\mkern-6mu/}% \displaystyle
    {/\mkern-6mu/}% \textstyle
    {/\mkern-5mu/}% \scriptstyle
    {/\mkern-5mu/}}}% \scriptscriptstyle

\newcommand\given[1][]{\:#1\vert\:}
\newcommand{\norm}[1]{\left\lVert#1\right\rVert}

\begin{document}

\title[Article Title]{FELRec: Efficient Handling of Item Cold-Start With Dynamic Representation Learning in Recommender Systems}


\author*[1]{\fnm{Kuba} \sur{Weimann}}\email{kuba.weimann@zib.de}

\author[1]{\fnm{Tim O.~F.} \sur{Conrad}}\email{conrad@zib.de}

\affil[1]{\orgdiv{Visual and Data-Centric Computing}, \orgname{Zuse Institute Berlin}, \orgaddress{\street{Takustr. 7}, \postcode{14195}, \state{Berlin}, \country{Germany}}}


\abstract{Recommender systems suffer from the cold-start problem whenever a new user joins the platform or a new item is added to the catalog. To address item cold-start, we propose to replace the embedding layer in sequential recommenders with a dynamic storage that has no learnable weights and can keep an arbitrary number of representations. In this paper, we present \texttt{FELRec}, a large embedding network that refines the existing representations of users and items in a recursive manner, as new information becomes available. In contrast to similar approaches, our model represents new users and items without side information and time-consuming finetuning, instead it runs a single forward pass over a sequence of existing representations. During item cold-start, our method outperforms similar method by $29.50\%$--$47.45$\%. Further, our proposed model generalizes well to previously unseen datasets in zero-shot settings. The source code is publicly available at \href{https://github.com/kweimann/FELRec}{https://github.com/kweimann/FELRec}.}

\keywords{Recommender systems, representation learning, cold-start, deep learning.}



\maketitle

\section{Introduction}
\label{sec:introduction}

One of the main goals of \textit{recommender systems} is to engage users with particular products by predicting their interests and suggesting appropriate items. In applications such as E-commerce, streaming platforms, or advertising, increasing user engagement ideally translates to higher profits for the respective business. The actual recommendation task is complex and difficult because recommender systems operate in inherently dynamic environments: new users join the system, existing users change their behavior, and new items are added to the catalog. For instance, users of a video streaming platform have various preferences that emerge over time as they consume more content throughout their streaming sessions. In such ever-changing environments, recommender systems must constantly adapt not only to new users but also to newly added items. Thus, one of the big challenges for recommender systems is the phase when no sufficient information is available to make good recommendations for a user. This is commonly called the \textit{cold-start} problem.

The cold-start problem is evident in classical approaches for modelling dynamic user behaviour, which are often based on sequential models. Sequential recommenders capture sequential patterns in the history of user activity. They rely on a sequence of user-item interactions to predict the next item that a user is likely to interact with. However, in the absence of interactions---during cold-start---the performance of sequential recommenders deteriorates. This is particularly visible in online systems, where new users and items arrive continuously \cite{vartak2017meta}. Various sequential models have been suggested in this area, such as Markov chains \cite{rendle2010factorizing, he2016fusing}, recurrent neural networks \cite{hidasi2016sessionbased, hidasi2018recurrent}, convolutional neural networks \cite{tang2018personalized, he2020lightgcn}, graph neural networks \cite{hamilton2017inductive, velivckovic2017graph, wang2019neural}, or attention-based networks \cite{kang2018selfattentive}. However, the main problems with cold-start have not been fully solved, and various further advancements have been proposed, including using available side information, meta-learning and finetuning (retraining).

Using available side information has been shown to mitigate cold-start problems. Side information can be derived from available user profiles, item attributes \cite{mooney2000content, saveski2014item} or even knowledge from 3rd party sources (e.g., Wikipedia \cite{narducci2016concept}). Side information can be combined with meta-learning approaches that estimate the preferences of users based only on a few user-item interactions \cite{vartak2017meta, lee2019melu, pan2019warm, sun2021form}. However, many meta-learning approaches often only address user cold-start, or rely on side information, which may be incomplete or missing due to privacy reasons. Zheng et al. \cite{zheng2021cold} propose a meta-learning framework to address item cold-start without side information, but their approach requires pretrained embeddings of existing items with many interactions to bootstrap the model. With time, these pretrained embeddings are likely to lose relevance, requiring retraining of the model to improve performance.

Another approach to deal with the item cold-start problem is finetuning---or retraining---the model to represent new items. However, finetuning is computationally expensive, and therefore often not feasible in online systems where latency matters. Addressing this problem, Kumar et al. \cite{kumar2019predicting} propose mutually recursive recurrent neural networks to generate dynamic embeddings of users and items, but they rely on the gradient of the loss function to update the model during test time. The downside is that this introduces a complex optimization process to keep the model up-to-date and significantly increases the computational effort. 

\begin{figure*}%
    \vskip 0.2in
    \centering
    \subfloat[\texttt{FELRec-Q}]{{\includegraphics[width=0.99\columnwidth]{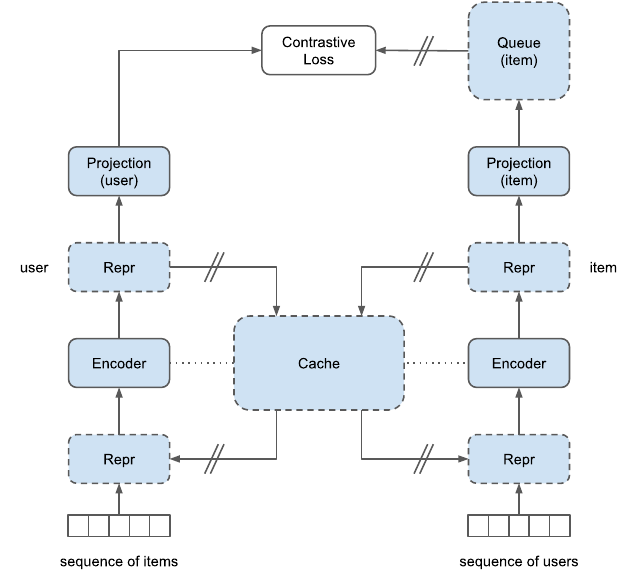} }}%
    \qquad
    \subfloat[\texttt{FELRec-P}]{{\includegraphics[width=0.99\columnwidth]{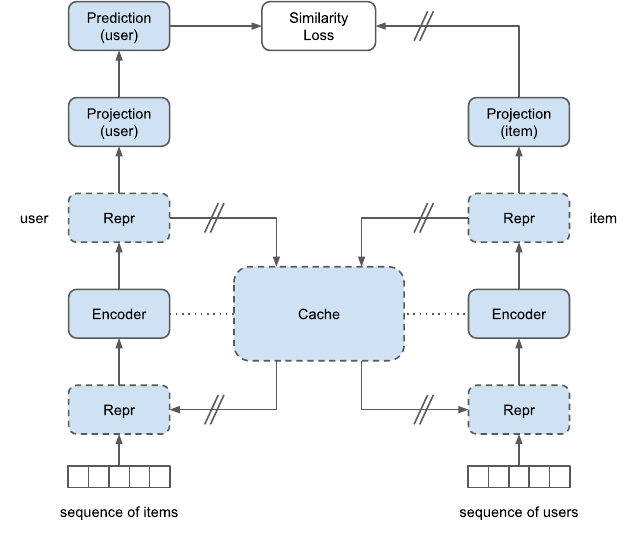} }}%
    \caption{Architectures of \texttt{FELRec-Q} (a) and \texttt{FELRec-P} (b). In a user-item interaction, the user is represented as a sequence of items, and the item as a sequence of users. First, \texttt{FELRec} fetches the vector representations of items and users from the cache. Next, our model encodes the sequences into new vector representations of the user and the item, which are then stored in the cache. \texttt{FELRec} is trained to make the representations of user and item similar using contrastive learning (a) adopted from \texttt{MoCo} \cite{he2020momentum}, or a similarity loss (b) adopted from \texttt{BYOL} \cite{grill2020bootstrap}. $\git$ means a stop-gradient operation.}%
    \label{fig:FELRec-architecture}%
    \vskip -0.2in
\end{figure*}

In this work, we take a different approach to item cold-start and propose a new paradigm for training sequential recommenders to address this issue. We develop a model that is capable of dynamically updating both user and item representations without any gradient updates, even in the absence of additional side information. We achieve this by replacing the traditional trained item embedding with a dynamic storage, which we call the \textit{cache}. This cache is not part of the learned model, therefore, it can expand flexibly to include new (or updated) items and users. We train our recommender system to interact with the cache using two distinct representation learning approaches: (1) contrastive learning \cite{hadsell2006contrastive} adopted from \textit{Momentum Contrast} (\texttt{MoCo}) \cite{he2020momentum}, and (2) the \textit{Bootstrap Your Own Latent} (\texttt{BYOL}) \cite{grill2020bootstrap} framework. Specifically, we leverage these representation learning techniques to teach the model how to update user and item representations through a single forward pass over a sequence of existing representations from the cache, ensuring similarity in representations when a user has interacted with an item. Figure \ref{fig:FELRec-architecture} illustrates our approach.

The major contributions of our work are as follows.

\begin{enumerate}
    \item By replacing the item embedding layer with a dynamic storage we enabled sequential recommenders to efficiently update and represent new items through a single forward pass, eliminating the need for gradient computations and side information. In item cold-start scenarios, our proposed method outperforms the strongest baseline by $29.50\%$--$47.45$\%.

    \item We showcase the adaptability of our proposed method, which in contrast to the traditional sequential recommenders generalizes to previously unseen recommendation datasets without any gradient updates (i.e., no finetuning is necessary) in zero-shot transfer scenarios.

    \item Thanks to the parameter-free cache, our proposed method requires a constant number of trainable parameters in contrast to the traditional sequential recommenders with a learned item embedding whose size depends on the number of items in the dataset.
\end{enumerate}

%%%%%%%%%%%%%%%%%%%%%%%%%%%%%%%%%%%%%%%%%%%%%%%%%%%%%%%%%%%%%%%%%%%%%%%%%%%%%%%%%%%%%%%%%%%%%%%%%%%%%%%%%
\section{Related work}
%%%%%%%%%%%%%%%%%%%%%%%%%%%%%%%%%%%%%%%%%%%%%%%%%%%%%%%%%%%%%%%%%%%%%%%%%%%%%%%%%%%%%%%%%%%%%%%%%%%%%%%%%
\label{sec:related}

We briefly explain why we choose to improve sequential recommenders in cold-start scenarios; how our model differentiates from other approaches that are dedicated to cold-start; and how we employ representation learning in comparison to other recommender systems.

\bmhead{Sequential recommendation}

The results of the Netflix Prize \cite{bennett2007netflix} demonstrated the effectiveness of matrix factorization (MF) \cite{mnih2008probabilistic, koren2009mf, rendle2009bpr} in solving recommendation tasks. In the subsequent years, several studies proposed to enhance MF with neural networks, e.g., by adding multi-layer perceptrons \cite{he2017neural} or auto-encoders \cite{sedhain2015autorec, wu2016collaborative}. Later, sequential recommenders began to replace MF due to better ability to capture patterns in user behaviour over time. The rich family of sequential models include Markov chains \cite{rendle2010factorizing, he2016fusing}, recurrent neural networks \cite{hidasi2016sessionbased, hidasi2018recurrent}, convolutional neural networks \cite{tang2018personalized, he2020lightgcn}, graph neural networks \cite{hamilton2017inductive, velivckovic2017graph, wang2019neural}, or attention-based networks \cite{kang2018selfattentive, sun2019bert4rec}, which were inspired by the advances in language modelling \cite{vaswani2017attention, devlin2019bert}. We also use a self-attentive model (Transformer \cite{vaswani2017attention}) to encode a sequence of items into a representation of the user to predict the next item in the sequence. However, we replace the embedding layer with a dynamic storage to accommodate new items, which is how we deal with cold-start.

\bmhead{Cold-start recommendation}

In the absence of user-item interactions, models can use side information, e.g., user profiles, item attributes \cite{mooney2000content, saveski2014item}, to better represent new users and items. Methods specifically designed to address cold-start include decision-making strategies, e.g., contextual bandits \cite{li2010contextual}, and meta-learning approaches \cite{vartak2017meta, lee2019melu, pan2019warm, zheng2021cold, sun2021form}, which are often based on the model-agnostic meta-learning (MAML) \cite{finn2017model} algorithm. In meta-learning, a model is trained to adapt to new (recommendation) tasks using only a few examples. Another line of work explicitly trains neural networks for cold-start through dropout \cite{volkovs2017dropoutnet}. Further, \cite{wei2021contrastive} address cold-start by learning better item representations with contrastive learning. \cite{kumar2019predicting, xu2020inductive} train neural networks to represent previously unseen nodes in dynamic graphs, which can be seen as a generalization of cold-start recommendation. In particular, \cite{kumar2019predicting} uses a memory module that is conceptually similar to the cache in \texttt{FELRec}. They store static and dynamic embeddings of users and items in the memory, but they update the model with the gradient of the loss function during test time. In contrast to many proposed methods, we address cold-start without use of side information, and we do not finetune our model to represent new items or users.

\bmhead{Representation learning}

In recent years, contrastive methods \cite{hadsell2006contrastive} have proven useful in learning image representations to solve various computer vision tasks \cite{oord2019representation, he2020momentum, chen2020simple, chen2021exploring, radford2021learning, zhai2022scaling}. Contrastive methods are trained to bring representations of different views of the same image (positive pair) closer while spreading the representations of views from different images (negative pairs). In our models, we adopt \texttt{MoCo}'s \cite{he2020momentum} queue as a dictionary of encoded samples and \texttt{BYOL}'s \cite{grill2020bootstrap} approach to learning representations. However, we do not use a momentum encoder, so our model is perhaps architecturally closer to \texttt{SimSiam} \cite{chen2021exploring}. In recommender systems, contrastive learning can be used as a regularization term to infer better representations of users and items \cite{yao2021selfsupervised, xie2022contrastive, wu2021selfsupervised, wei2021contrastive, liu2021contrastive}, or to pretrain the representations for a later finetuning \cite{zhou2020s3rec, cheng2021learning}. Further, \cite{liu2021bcontrastive} introduce a debiased contrastive loss to address sample bias in negative sampling strategies such as Bayesian Personalized Ranking (BPR) \cite{rendle2009bpr}. In comparison to BPR, contrastive losses use many negative samples as opposed to just one, whereas \texttt{BYOL} gets rid of negative pairs altogether. We also train our model using only the contrastive (or similarity) loss.

%%%%%%%%%%%%%%%%%%%%%%%%%%%%%%%%%%%%%%%%%%%%%%%%%%%%%%%%%%%%%%%%%%%%%%%%%%%%%%%%%%%%%%%%%%%%%%%%%%%%%%%%%
\section{FELRec}
%%%%%%%%%%%%%%%%%%%%%%%%%%%%%%%%%%%%%%%%%%%%%%%%%%%%%%%%%%%%%%%%%%%%%%%%%%%%%%%%%%%%%%%%%%%%%%%%%%%%%%%%%
\label{sec:method}

In this section, we describe our recommender system called \texttt{FELRec}\footnote{\textbf{F}orget \textbf{E}mbedding \textbf{L}ayers in \textbf{Rec}ommenders}. The architecture of our model resembles that of traditional sequential recommenders, which typically consist of three components: (1) an embedding layer, (2) a sequence-to-sequence encoder, and (3) a prediction layer. However, unlike conventional models, the embedding layer in \texttt{FELRec} is replaced by a dynamic storage we call \textit{cache}.

The cache can store an arbitrary number of representations of users and items. It has no trainable weights; therefore, updating a representation simply involves replacing the corresponding entry in the cache, no finetuning or retraining is necessary. Using either a contrastive loss \cite{hadsell2006contrastive}, as adopted from \texttt{MoCo} \cite{he2020momentum}, or a similarity loss as adopted from \texttt{BYOL} \cite{grill2020bootstrap}, we train \texttt{FELRec} to interact with the cache---update user and item representations with a single forward pass over a sequence of existing representations from the cache---and simultaneously solve the recommendation task.

%%%%%%%%%%%%%%%%%%%%%%%%%%%%%%%%%%%%%%%%%%%%%%%%%%%%%%%%%%%%%%%%%%%%%%%%%%%%%%%%%%%%%%%%%%%%%%%%%%%%%%%%%
\subsection{Problem statement}

We begin by defining the primary objective of recommender systems. Let $\mathcal{U}$ denote the set of users, $\mathcal{X}$ the set of items and $S_u^t = \left[x_u^1, \cdots , x_u^t\right]$ a sequence of items $x_u^i \in \mathcal{X}$, with which the user $u \in \mathcal{U}$ has interacted until some (relative) timestamp $t$. Since we do not use any kind of additional side information, users and items do not have any features. Given a sequence $S_u^t$, the objective is to predict the next likely item $x_u^{t+1}$. In other words, we model the probability of a future user-item interaction over all available items $P(X = x_u^{t+1} \given S_u^t)$.

In scenarios involving item cold-start, where an item is newly introduced to the catalogue, the sequence of user interactions with this item, denoted as $S_x^t$, is typically short or non-existent. This lack of substantial interaction history presents a significant challenge in accurately predicting the item’s relevance and likelihood of being chosen by future users.

%%%%%%%%%%%%%%%%%%%%%%%%%%%%%%%%%%%%%%%%%%%%%%%%%%%%%%%%%%%%%%%%%%%%%%%%%%%%%%%%%%%%%%%%%%%%%%%%%%%%%%%%%
\subsection{Cache}

In traditional sequential recommenders, once an item embedding is learned, altering it requires either finetuning or retraining the entire model. This issue is exacerbated in situations lacking supplementary side information, where item embeddings depend solely on interaction data, rendering generic methods for content-based embedding (e.g., using item metadata) infeasible. Consequently, for new items with a short interaction history, the system often struggles to generate precise and informative embeddings, leading to suboptimal recommendations due to poorly defined or overly generic vector representations.

To address these limitations, our proposed model introduces a new component called \textit{cache}, a dynamic storage replacement for the conventional embedding layer. Unlike traditional embeddings, the cache does not require gradient computation to alter existing representations. It primarily functions as a look-up table for item and user representations. These representations, similar to those in standard sequential recommenders, are then fed into the sequential model for extracting user preferences for future items.

The cache can be implemented as any primitive collection such as an array or dictionary. Initially empty, the cache represents non-existent entries with zero vectors. As the cache is parameter-free and does not learn item representations in the traditional sense, it can expand flexibly to include new items and update existing item representations without necessitating complex gradient updates.

In our experimental setup, we impose no constraints on the cache size. Limiting its capacity could adversely affect performance, potentially leading to an exacerbated popularity bias, as less popular items might be disproportionately excluded \citep{cremonesi2010performance}. Instead, we manage the cache contents dynamically, enabling the transfer of data between storage and processing units as needed. This approach aligns with how online recommender systems operate, maintaining pre-computed user and item representations for efficient real-time recommendation delivery.

%%%%%%%%%%%%%%%%%%%%%%%%%%%%%%%%%%%%%%%%%%%%%%%%%%%%%%%%%%%%%%%%%%%%%%%%%%%%%%%%%%%%%%%%%%%%%%%%%%%%%%%%%
\subsection{Model architecture}

The cache serves an important role, enabling our model to dynamically include and update representations without gradient computations. However, this flexibility introduces a new challenge: the model must learn to (re-)compute these representations so they accurately reflect the evolving preferences of users, thereby maintaining their relevance for effective item recommendation. To overcome this challenge, we integrate two established representation learning methods: \textit{Momentum Contrast} (MoCo) \citep{he2020momentum} and \textit{Bootstrap Your Own Latent} (BYOL) \citep{grill2020bootstrap}, leading to the development of two model variants, \texttt{FELRec-Q} and \texttt{FELRec-P}, respectively. These adaptations enable our model to effectively interact with the cache, ensuring that it retrieves and updates representations in a manner that preserves similarity when a user interacts with an item.

The process of updating the cache with new user-item interactions begins with a new approach. Drawing inspiration from sequential recommender systems, \texttt{FELRec} characterizes a user by the sequence of their past item interactions and an item by the sequence of its user interactions, an essential step due to the absence of pre-learned item embeddings. This design positions the cache as a dynamic memory, encapsulating the entire user-item interaction graph, including both user and item representations. \texttt{FELRec} retrieves these vector representations from the cache and encodes them into updated vector representations for both the user and item, which are subsequently stored back in the cache. This approach ensures a seamless integration of new interactions into the model.

\subsubsection*{FELRec-Q}

The first variant, \texttt{FELRec-Q}, uses a contrastive loss \cite{hadsell2006contrastive} to make the vector representations of a user and an item similar. Specifically, we train the model for a look-up task in a dictionary of items. Consider an encoded user-item interaction $(\hat{u}, \hat{x}_{+})$ and a dictionary of encoded items $\left\{\hat{x}_0, \cdots, \hat{x}_{K-1}\right\}$ that contains $\hat{x}_{+}$. We use InfoNCE \cite{oord2019representation}, a contrastive loss function, to identify the positive sample $\hat{x}_{+}$ among negative samples in the dictionary:

\begin{equation}
\label{eq:FELRec-user-loss}
    \mathcal{L}_u = -\log\frac{\exp(\hat{u}\,\cdot\,\hat{x}_{+}\,/\,\tau)}{\sum_{i=0}^{K}\exp(\hat{u}\,\cdot\,\hat{x}_i\,/\,\tau)}
\end{equation}

where $\tau$ is a temperature hyperparameter \cite{wu2018unsupervised} and similarity is measured by dot product. Similar to \texttt{MoCo} \cite{he2020momentum}, we maintain the dictionary as a queue of encoded items. 

To formalize the model, let $(u, x)$ be a user-item interaction, $f_u$ user encoder, $f_x$ item encoder, and $g_u$, $g_x$ respective user and item projection networks, which are small multi-layer perceptrons (MLP). We consider $\hat{u}$ the projection of a user representation $\hat{u} = g_u(f_u(u))$. Similarly, $\hat{x}$ is the projection of an item representation $\hat{x} = g_x(f_x(x))$, which we refer to as $\hat{x}_{+}$ in equation \ref{eq:FELRec-user-loss}. We share parameters between the encoders such that $f_u = f_x$. 
During backpropagation, we do not update the weights of the item encoder $f_x$ because the queue contains data from numerous past minibatches, making it computationally infeasible to backpropagate through all these stored computations \cite{he2020momentum}. Furthermore, we do not update the cache with the gradient of the loss function. The architecture of \texttt{FELRec-Q} is illustrated in Figure \ref{fig:FELRec-architecture}a.

We symmetrize the loss by simultaneously training the model to identify the positive user among negative users in a separate dictionary $\left\{\hat{u}_0, \cdots, \hat{u}_{K-1}\right\}$:

\begin{equation}
    \mathcal{L}_x = -\log\frac{\exp(\hat{x}\,\cdot\,\hat{u}_{+}\,/\,\tau)}{\sum_{i=0}^{K}\exp(\hat{x}\,\cdot\,\hat{u}_i\,/\,\tau)}
\end{equation}

where positive user $\hat{u}_{+} = \hat{u}$. Here, we do not update the weights of user encoder $f_u$. The final loss is the sum of these two losses: $\mathcal{L} = \mathcal{L}_{u} + \mathcal{L}_{x}$.

\subsubsection*{FELRec-P}

In addition to the above presented model architecture, we implement a second variant of \texttt{FELRec} that uses a different representation learning paradigm adopted from \texttt{BYOL} \cite{grill2020bootstrap}. We call this variant \texttt{FELRec-P}. Designed for image representation learning, \texttt{BYOL} minimizes a similarity loss between representations of different augmented views of the same image. We adjust \texttt{BYOL} to minimize a similarity loss between representations of user and item that interact with each other. In accordance with \texttt{BYOL}, we define two further MLPs: user prediction network $h_u$ and item prediction network $h_x$. We train the model by minimizing the mean squared error between the normalized user prediction $h_u(\hat{u})$ and the item projection $\hat{x}$:

\begin{equation}
    \mathcal{L}_u = 2 - 2 \,\cdot\, \frac{h_u(\hat{u})\,\cdot\,\hat{x}}{\norm{h_u(\hat{u})}_2 \,\cdot\,\norm{\hat{x}}_2}\,\cdot
\end{equation}

Following the original approach, we stop the gradient in the item branch to prevent a collapse of the model into trivial solutions and stabilize the training process \cite{grill2020bootstrap}. Further, we symmetrize the loss by computing the mean squared error between the normalized item prediction and user projection:

\begin{equation}
    \mathcal{L}_x = 2 - 2 \,\cdot\, \frac{h_x(\hat{x})\,\cdot\,\hat{u}}{\norm{h_x(\hat{x})}_2 \,\cdot\,\norm{\hat{u}}_2}\,\cdot
\end{equation}

Again, we stop the gradient in the user branch. The final loss is the sum of these two losses: $\mathcal{L} = \mathcal{L}_{u} + \mathcal{L}_{x}$. The gradient stopping and reuse of the encoder make the model a \textit{siamese neural network} that is conceptually similar to \texttt{SimSiam} \cite{chen2021exploring}. Figure \ref{fig:FELRec-architecture}b illustrates the architecture of \texttt{FELRec-P}.

%%%%%%%%%%%%%%%%%%%%%%%%%%%%%%%%%%%%%%%%%%%%%%%%%%%%%%%%%%%%%%%%%%%%%%%%%%%%%%%%%%%%%%%%%%%%%%%%%%%%%%%%%
\subsection{Encoder}

The sequence-to-sequence encoder in \texttt{FELRec} is a Transformer \cite{vaswani2017attention} followed by a pooling layer. We use only the encoder part of the Transformer with the following hyperparameters: $N=3$ layers, $d_{model}=128$ output dimensions (model dimension), $d_{ff}=256$ inner dimensions, $h=4$ attention heads, $P_{drop}=0.1$ dropout rate. The self-attention mechanism \cite{vaswani2017attention} in our encoder is bidirectional, i.e., every position in the sequence attends to the context to its left (past) and right (future). We mask only the empty positions in the input sequence. The pooling layer averages out all encodings, which are not masked, to form a single pooled vector. Finally, an additional linear layer transforms the pooled vector into a new vector representation of a user or item.

To better distinguish between input sequences, we add a \textit{type embedding} to every position in the sequence. The type embedding is a learned embedding that symbolizes either the user or an item. It has the same dimension as the model.

%%%%%%%%%%%%%%%%%%%%%%%%%%%%%%%%%%%%%%%%%%%%%%%%%%%%%%%%%%%%%%%%%%%%%%%%%%%%%%%%%%%%%%%%%%%%%%%%%%%%%%%%%
\subsection{Projection \& prediction networks}

The vector representations of users and items are projected using small multi-layer perceptrons (MLP). The projection network in \texttt{FELRec-Q} consists of a hidden layer followed by batch normalization \cite{ioffe2015batch}, rectified linear units (ReLU) \cite{nair2010rectified} and a linear output layer. The output size of the feedforward layers is the same as the model. \texttt{FELRec-P} utilizes a bottleneck structure in the projection and prediction networks, similar to \texttt{BYOL} \cite{grill2020bootstrap} and \texttt{SimCLR} \cite{chen2020simple}. The MLPs follow the same architecture as the projection network in \texttt{FELRec-Q}, but the dimension of the hidden layer is double the model dimension ($256$) and the dimension of the output is half of the model dimension ($64$).

%%%%%%%%%%%%%%%%%%%%%%%%%%%%%%%%%%%%%%%%%%%%%%%%%%%%%%%%%%%%%%%%%%%%%%%%%%%%%%%%%%%%%%%%%%%%%%%%%%%%%%%%%
\subsection{Model training}

We train both variants of \texttt{FELRec} using stochastic gradient descent (SGD) with cosine learning rate schedule \cite{loshchilov2016sgdr}, without restarts, over $100$ epochs, with a warm-up period of $10$ epochs. The base learning rate is $0.01$, the momentum parameter is $0.9$ and the batch size is $1024$. We use no weight decay. The temperature hyperparameter in \texttt{FELRec-Q} is $0.07$ and the queue size is $8192$. The input sequences consist of up to $64$ of the most recent items or users. 

A distinctive aspect of our training methodology involves the cache management. At the onset of each epoch, the cache is emptied. This intentional clearing of the cache is a strategic measure designed to force the model to consistently learn from scratch in computing new representations, rather than merely refining existing ones. Moreover, in alignment with the operational logic of online systems, the dataset is not shuffled. Instead, like in an online system, we follow the order of user-item interactions, which is given by the timestamps. This way, we avoid introducing information about the future while computing new vector representations of users and items. 

%%%%%%%%%%%%%%%%%%%%%%%%%%%%%%%%%%%%%%%%%%%%%%%%%%%%%%%%%%%%%%%%%%%%%%%%%%%%%%%%%%%%%%%%%%%%%%%%%%%%%%%%%
\subsection{Model evaluation}

When evaluating the model, we keep the existing vector representations in the cache but discard the projection (and prediction) networks. We iterate once over the test data following the order of user-item interactions. We compute scores (logits) by directly comparing the vector representations of users and items with cosine similarity. This method of evaluation resembles conventional sequential recommenders whose embedding and prediction layers share weights and their dot product returns logits. In contrast to the conventional models, we use representations from the cache instead of embedding weights.

A distinctive feature of \texttt{FELRec}, setting it apart from these traditional models, is its dynamic nature during the inference stage. As new information becomes available---manifested through each new user-item interaction---the model updates the corresponding representations with just a single forward pass. This process is a significant deviation from conventional models, where representations are static during inference.

%%%%%%%%%%%%%%%%%%%%%%%%%%%%%%%%%%%%%%%%%%%%%%%%%%%%%%%%%%%%%%%%%%%%%%%%%%%%%%%%%%%%%%%%%%%%%%%%%%%%%%%%%
\section{Experimental setup}
%%%%%%%%%%%%%%%%%%%%%%%%%%%%%%%%%%%%%%%%%%%%%%%%%%%%%%%%%%%%%%%%%%%%%%%%%%%%%%%%%%%%%%%%%%%%%%%%%%%%%%%%%
\label{sec:experiments}

This section describes the data and the experimental setup that highlights the advantages and downsides of our model in various recommendation scenarios. Oftentimes, studies similar to ours only use small to medium-sized datasets in their evaluation. We train and evaluate \texttt{FELRec} on large datasets containing $25$ million user-item interactions. We compare our model with one strong baseline method for item cold-start, and four additional baseline methods that excel at recommending items that are not new. We evaluate all models with two metrics that differ in how they select the catalog of items to recommend from.

%%%%%%%%%%%%%%%%%%%%%%%%%%%%%%%%%%%%%%%%%%%%%%%%%%%%%%%%%%%%%%%%%%%%%%%%%%%%%%%%%%%%%%%%%%%%%%%%%%%%%%%%%
\subsection{Datasets}

We use the following two datasets to train our model, and subsequently to evaluate it.

\begin{itemize}
    \item \textit{MovieLens} is a benchmark dataset of movie ratings. We use the MovieLens 25M Dataset \cite{harper2015movielens}, which includes $162,541$ users, $59,047$ movies and $25$ million movie ratings. We treat every rating as a user-item interaction.
    \item \textit{Twitch} is a dataset of users watching streams on Twitch collected over $43$ days in $10$-minute intervals \cite{rappaz2021twitch}. We treat every record as a user-item interaction. We sample $25$ million interactions, which include $817,625$ users and $432,813$ streams. We ignore the total time spent on a stream. 
\end{itemize}

In all experiments, we assess how well models generalize to the future. We sort user-item interactions in chronological order based on the timestamps and treat the first $80\%$ of interactions as training set, the following $10\%$ as validation set, and the remaining $10\%$ as test set. We partition the test set to give a more holistic overview of the performance of our method in different recommendation scenarios, not only item cold-start. 

\begin{itemize}
    \item \textit{Observed} includes user-item interactions where both user and item are part of the training set.
    \item \textit{New users} includes user-item interactions where item is part of the training set, but user is not.
    \item \textit{New items} includes user-item interactions where item is not part of the training set, and user may or may not be part of the training set.
\end{itemize}

The partitioning also highlights which models can represent new users or new items. For instance, sequential recommenders cannot recommend items whose embedding has not been learned, so we cannot evaluate them on the new items partition. We also report the \textit{total} performance of models on all partitions that they can be evaluated on. Note that these partitions are different among methods, so the total performances are not always comparable.

The effective test set is larger for \texttt{FELRec} and methods that can represent new items. During test time, these models make recommendations from a larger pool of items because the pool also includes new items that appear after the training phase.

\begin{table*}[t]
\caption{Recommendation performance on MovieLens 25M. The percentages below the test partitions indicate the proportion of all test data. Lower rank is better; higher HR@10 is better. (See Section \ref{sec:evaluation} for details.)}
\label{table-ml-25m}
\vskip 0.15in
\begin{center}
\begin{small}
\begin{sc}

\begin{tabular}{llcccccc|cc}
\toprule
\multirow{2}{*}{model} & \multicolumn{1}{c}{\multirow{2}{*}{params}} & \multicolumn{2}{c}{\begin{tabular}[c]{@{}c@{}}observed\\ (9.47\%)\end{tabular}} & \multicolumn{2}{c}{\begin{tabular}[c]{@{}c@{}}new users\\ (77.77\%)\end{tabular}} & \multicolumn{2}{c}{\begin{tabular}[c]{@{}c@{}}new items\\ (12.76\%)\end{tabular}} & \multicolumn{2}{|c}{total} \\
                       & \multicolumn{1}{c}{}                        & \multicolumn{1}{l}{rank}               & \multicolumn{1}{l}{hr@10}              & \multicolumn{1}{l}{rank}                & \multicolumn{1}{l}{hr@10}               & \multicolumn{1}{l}{rank}                & \multicolumn{1}{l}{hr@10}               & \multicolumn{1}{|l}{rank}              & \multicolumn{1}{l}{hr@10}              \\
\midrule
\texttt{BPR-MF}                 & 22.06M                  & 0.119                                 & 68.16\%                                 & ---                                      & ---                                        & ---                                      & ---                                        & 0.119      & 68.16\%      \\
\texttt{GRU4Rec}                & 17.93M                  & 0.135                                 & 71.02\%                                 & 0.054                                  & 89.46\%                                  & ---                                      & ---                                        & 0.063      & 87.46\%      \\
\texttt{GRU4Rec-BPTT}           & 4.72M                   & 0.110                                      & 73.40\%                                        & 0.041                                       & 90.74\%                                         & ---                                       & ---                                         & 0.049           & 88.85\%             \\
\texttt{SASRec}                 & 4.82M                  & \textbf{0.075}                                 & \textbf{80.17}\%                                 & \textbf{0.026}                                  & \textbf{93.88}\%                                  & ---                                      & ---                                        & 0.031      & 92.39\%      \\
\texttt{JODIE*}                 & 28.56M                  & 0.137                                 & 67.94\%                                 & 0.067                                  & 83.14\%                                  & 0.186                                      & 57.66\%                                        & 0.089      & 78.45\%      \\
\midrule
\texttt{FELRec-Q}                & 0.48M                   & 0.095                                 & 74.79\%                                 & 0.038                                  & 90.81\%                                  & \textbf{0.102}                                  & \textbf{79.39}\%                                  & 0.052      & 87.85\%      \\
\texttt{FELRec-P}                 & 0.58M                   & 0.086                                 & 75.76\%                                 & 0.038                                  & 90.88\%                                  & 0.139                                  & 73.49\%                                  & 0.055      & 87.23\%      \\
\texttt{FELRec-Q-reset}          & 0.48M                   & 0.102                                 & 65.08\%                                 & 0.070                                  & 77.58\%                                  & 0.203                                  & 51.73\%                                  & 0.090      & 73.09\%      \\
\texttt{FELRec-Q-zero}       & 0.48M                   & 0.128                                 & 56.17\%                                 & 0.077                                  & 75.56\%                                  & 0.206                                  & 48.34\%                                  & 0.098      & 70.25\%    \\                  
\bottomrule
\end{tabular}

\end{sc}
\end{small}
\end{center}
\vskip -0.1in
\end{table*}

%%%%%%%%%%%%%%%%%%%%%%%%%%%%%%%%%%%%%%%%%%%%%%%%%%%%%%%%%%%%%%%%%%%%%%%%%%%%%%%%%%%%%%%%%%%%%%%%%%%%%%%%%
\subsection{Baseline methods}

We select five established baseline methods, where each is representative of a particular category of data it can be evaluated on (i.e., test partition). \texttt{JODIE*} is a strong baseline method similar to our model---it can represent new items without the use of side information. \texttt{GRU4Rec}, \texttt{GRU4Rec-BPTT} and \texttt{SASRec} are recognized sequential recommenders that can represent new users as a sequence of items, but lack a mechanism to represent new items. Finally, \texttt{BPR-MF} is a well-known factor model that cannot represent users or items if they were not observed during training. \texttt{SASRec}, \texttt{GRU4Rec}, \texttt{GRU4Rec-BPTT} and \texttt{BPR-MF} can be considered examples of retraining a model at the time when training data ends. 

\begin{itemize}
    \item \texttt{BPR-MF} \cite{rendle2009bpr} is a matrix factorization model that is optimized with respect to the Bayesian Personalized Ranking (BPR) criterion.
    \item \texttt{GRU4Rec} \cite{hidasi2016sessionbased} is a sequential recommender based on gated recurrent units (GRU) \cite{cho2014learning} for session-based recommendations.
    \item \texttt{GRU4Rec-BPTT} is our alternative implementation of \texttt{GRU4Rec} that uses backpropagation through time (BPTT) \cite{werbos1990backprop} to train the recurrent neural network.
    \item \texttt{SASRec} \cite{kang2018selfattentive} is a sequential recommender based on the self-attention mechanism \cite{vaswani2017attention}. We use the same encoder in our \texttt{FELRec}.
    \item \texttt{JODIE*} \cite{kumar2019predicting} is a recurrent neural network that can estimate future embedding trajectories of users and items. Similar to our model, \texttt{JODIE*} uses a memory module to represent new users and new items, but during test time, \texttt{JODIE*} is updated with the gradient of the loss function. We mark the model with an asterisk (\texttt{*}) because we made significant changes to the model to circumvent problems with memory consumption on large datasets.
\end{itemize}

Some baseline methods needed adjustments to adapt them to our recommendation tasks. We discuss the changes to their original implementations in Appendix \ref{app:baselines}.

%%%%%%%%%%%%%%%%%%%%%%%%%%%%%%%%%%%%%%%%%%%%%%%%%%%%%%%%%%%%%%%%%%%%%%%%%%%%%%%%%%%%%%%%%%%%%%%%%%%%%%%%%
\subsection{Evaluation}
\label{sec:evaluation}

We evaluate the performance of all models using the following two metrics.

\begin{itemize}
    \item \textit{Normalized rank} (\textit{rank}) is the position of the item picked by the user (i.e., ground-truth item) from the catalog of all recommendations. This list of recommended items is sorted according to the calculated similarity measure. The higher the ground-truth item on that list, the better the recommendation matches the interests of the user.
    
    If a model can only recommend learned items, the catalog will never change at test time. However, \texttt{FELRec} can compute representations of new items at test time, so the list will grow over time. Therefore, we normalize the ranks by dividing them by the size of the list at a given timestamp. The final score is the mean over all normalized ranks. It is a value between $0$ and $1$. The lower this value, the better the performance.
    \item \textit{Hit rate@10} (\textit{HR@10}) counts the percentage of recommendations where the item picked by the user (i.e., ground-truth item) is among the top $10$ recommended items in the catalog. To calculate HR@10, we draw 100 negative samples (items) for every interaction.
\end{itemize}

Briefly put: rank measures recommendation performance using the entire catalog of items, whereas HR@10 uses only a subset of the catalog.

After every epoch, we validate each model using the rank metric. At the end of the training, we revert the models to a checkpoint with the best validation rank. Then, we report the rank and HR@10 metric on the three disjoint partitions of the test set.

\begin{table*}[t]
\caption{Recommendation performance on Twitch. The percentages below the test partitions indicate the proportion of all test data. Lower rank is better; higher HR@10 is better. (See Section \ref{sec:evaluation} for details.)}
\label{table-twitch}
\vskip 0.15in
\begin{center}
\begin{small}
\begin{sc}
\begin{tabular}{llcccccc|cc}
\toprule

\multicolumn{1}{l}{\multirow{2}{*}{model}} & \multirow{2}{*}{params} & \multicolumn{2}{c}{\begin{tabular}[c]{@{}c@{}}observed\\ (93.18\%)\end{tabular}} & \multicolumn{2}{c}{\begin{tabular}[c]{@{}c@{}}new users\\ (3.42\%)\end{tabular}} & \multicolumn{2}{c}{\begin{tabular}[c]{@{}c@{}}new items\\ (3.40\%)\end{tabular}} & \multicolumn{2}{|c}{total}                            \\
\multicolumn{1}{r}{}                       &                         & \multicolumn{1}{l}{rank}               & \multicolumn{1}{l}{hr@10}               & \multicolumn{1}{l}{rank}               & \multicolumn{1}{l}{hr@10}               & \multicolumn{1}{l}{rank}               & \multicolumn{1}{l}{hr@10}               & \multicolumn{1}{|l}{rank} & \multicolumn{1}{l}{hr@10} \\
\midrule
\texttt{BPR-MF}                 & 152.33M                 & 0.011                                  & 97.29\%                                 & ---                                      & ---                                       & ---                                      & ---                                       & 0.011      & 97.29\%      \\
\texttt{GRU4Rec}                & 199.19M                 & 0.032                                       & 92.02\%                                        & 0.065                                       & 82.79\%                                        & ---                                       & ---                                        & 0.033           & 91.69\%             \\
\texttt{GRU4Rec-BPTT}           & 49.95M                  & 0.039                                       & 91.71\%                                        & 0.090                                       & 80.43\%                                        & ---                                       & ---                                        & 0.041           & 91.31\%             \\
\texttt{SASRec}                 & 50.05M                  & \textbf{0.006}                                  & \textbf{98.43}\%                                 & \textbf{0.026}                                  & \textbf{93.47}\%                                 & ---                                      & ---                                       & 0.007      & 98.26\%      \\
\texttt{JODIE*}                 & 160.25M                  & 0.059                                  & 86.73\%                                 & 0.144                                  & 66.62\%                                 & 0.314                                      & 40.92\%                                       & 0.070      & 84.49\%      \\
\midrule
\texttt{FELRec-Q}                & 0.48M                   & 0.021                                  & 95.13\%                                 & 0.084                                  & 84.59\%                                 & \textbf{0.165}                                  & \textbf{52.99}\%                                 & 0.027      & 93.34\%      \\
\texttt{FELRec-P}                 & 0.58M                   & 0.019                                  & 94.60\%                                 & 0.068                                  & 80.58\%                                 & 0.181                                  & 49.11\%                                 & 0.027      & 92.58\%      \\
\texttt{FELRec-Q-reset}          & 0.48M                   & 0.039                                  & 88.45\%                                 & 0.156                                  & 61.00\%                                 & 0.385                                  & 43.35\%                                 & 0.055      & 85.98\%      \\
\texttt{FELRec-Q-zero}       & 0.48M                    & 0.038                                  & 91.00\%                                 & 0.103                                  & 71.72\%                                 & 0.401                                  & 41.08\%                                 & 0.053      & 88.65\%                  \\
\bottomrule
\end{tabular}
\end{sc}
\end{small}
\end{center}
\vskip -0.1in
\end{table*}

%%%%%%%%%%%%%%%%%%%%%%%%%%%%%%%%%%%%%%%%%%%%%%%%%%%%%%%%%%%%%%%%%%%%%%%%%%%%%%%%%%%%%%%%%%%%%%%%%%%%%%%%%
\section{Results}
%%%%%%%%%%%%%%%%%%%%%%%%%%%%%%%%%%%%%%%%%%%%%%%%%%%%%%%%%%%%%%%%%%%%%%%%%%%%%%%%%%%%%%%%%%%%%%%%%%%%%%%%%
\label{sec:results}

In this section, we present and analyze the results of the experiments described in the previous section for the various recommendation scenarios. We show that \texttt{FELRec} outperforms \texttt{JODIE*} during item cold-start and beyond, but it falls behind on existing items compared to strong sequential recommenders such as \texttt{SASRec}. The results demonstrate that the quality of recommendations by \texttt{FELRec} significantly improves over time, necessitating large datasets for training. Further, in contrast to other models, \texttt{FELRec} generalizes well to unseen datasets without finetuning, in what is called zero-shot transfer to downstream tasks. We also discuss model complexity and popularity bias, and we highlight critical parts of the model in ablation studies. Finally, we show how to easily find users with similar interests using their representations.

%%%%%%%%%%%%%%%%%%%%%%%%%%%%%%%%%%%%%%%%%%%%%%%%%%%%%%%%%%%%%%%%%%%%%%%%%%%%%%%%%%%%%%%%%%%%%%%%%%%%%%%%%
\subsection{Recommendation of cold-start items}

We begin by comparing the recommendation performance of \texttt{FELRec} and \texttt{JODIE*} with a focus on the item cold-start phase. \texttt{FELRec} outperforms \texttt{JODIE*} on all test partitions across both datasets. During item cold-start, the difference lies between $29.50\%$ and $47.45$\%, depending on the dataset and metric. On MovieLens 25M (Table \ref{table-ml-25m}), the performance of \texttt{FELRec-Q} on new items is better by $0.084$ rank and $21.73\%$ HR@10. On Twitch (Table \ref{table-twitch}), the performance is better by $0.149$ rank and $12.07\%$ HR@10. 

The performances of \texttt{FELRec-Q} and \texttt{FELRec-P} are very similar across all test partitions, but \texttt{FELRec-Q} handles the recommendation of new items better. Therefore, we concentrate on \texttt{FELRec-Q} in the following experiments.

%%%%%%%%%%%%%%%%%%%%%%%%%%%%%%%%%%%%%%%%%%%%%%%%%%%%%%%%%%%%%%%%%%%%%%%%%%%%%%%%%%%%%%%%%%%%%%%%%%%%%%%%%
\subsection{Recommendation of existing items}

We compare \texttt{FELRec} with the sequential recommender that makes the best recommendations of existing items. Among the baseline methods, \texttt{SASRec} achieves the best results on the observed and new users partitions. On MovieLens 25M (Table \ref{table-ml-25m}), \texttt{SASRec} outperforms \texttt{FELRec-Q} by $0.02$ rank ($5.38\%$ HR@10) and $0.012$ rank ($3.07\%$ HR@10), respectively. On Twitch (Table \ref{table-twitch}), the performance of \texttt{SASRec} is better by $0.015$ rank ($3.30\%$ HR@10) and $0.058$ rank ($8.88\%$ HR@10), respectively. 

\texttt{FELRec}'s relatively lower performance can primarily be traced back to its distinct training strategy. Contrary to traditional sequential recommendation systems, which progressively refine their embeddings across multiple training epochs, \texttt{FELRec} resets its cache at the beginning of each epoch. This frequent resetting disrupts the continuity of the learning process, preventing the model from developing and retaining increasingly refined item and user representations compared to models that maintain and update their embeddings consistently over time.

Furthermore, we suspect that the lower performance of \texttt{FELRec} is in part related to the distribution of interactions among the partitions. On Twitch, the majority of interactions occur between observed users and items whose representations can be learned during training. Therefore, models with a learned embedding, even architecturally as simple as \texttt{BPR-MF}, can achieve excellent performance. On MovieLens 25M, where most interactions come from new users, models are incentivized to learn how to represent new users well.

The results suggest that \texttt{SASRec} is a better choice in a conventional setting where only items already observed during training are recommended. However, our method compensates for the lower performance with its ability to represent new, previously unobserved, items during inference. 

\begin{figure*}[t]%
    \centering
    \subfloat[MovieLens 25M]{{\includegraphics[width=0.9\columnwidth]{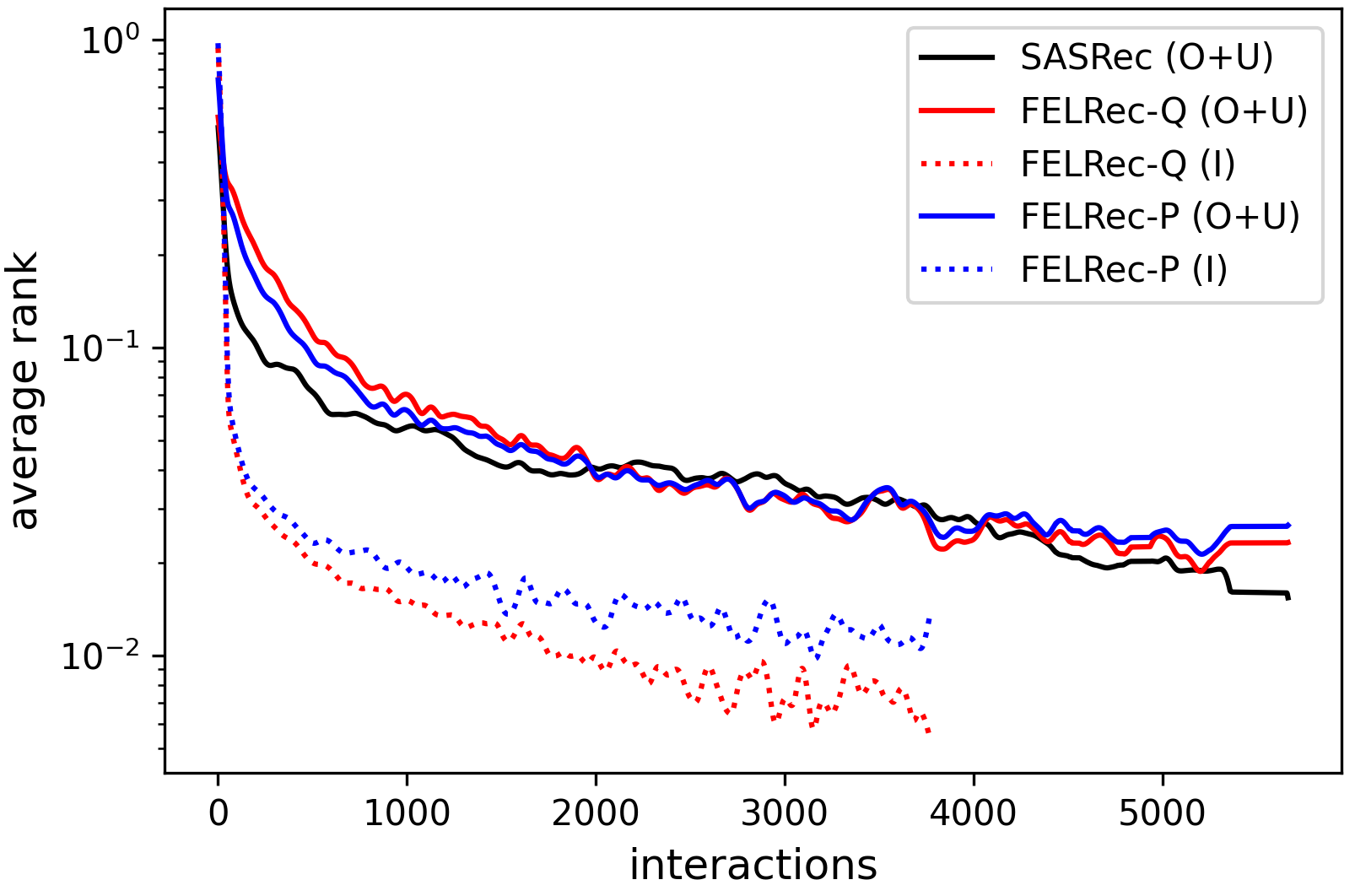} }}%
    \qquad
    \subfloat[Twitch]{{\includegraphics[width=0.9\columnwidth]{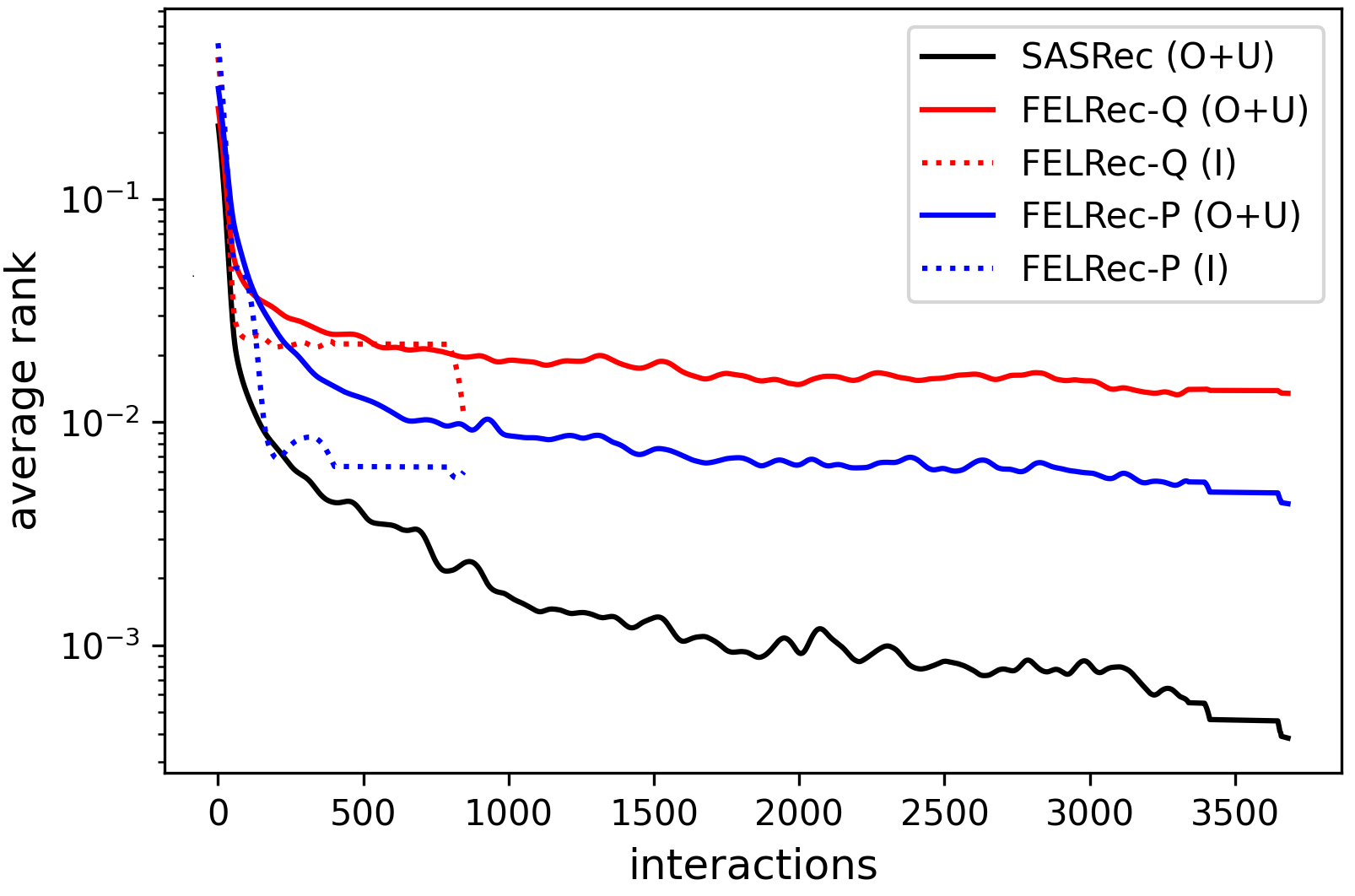} }}%
    \caption{Recommendation performance of \texttt{FELRec} and \texttt{SASRec} on MovieLens 25M (a) and Twitch (b) depending on the number of users that have interacted with the ground-truth item. Lower average rank is better. "I" is the new items partition, "O+U" are the observed and new users partitions. Performance improves over time.}%
    \label{fig:item-rank}%
    \vskip -0.2in
\end{figure*}

%%%%%%%%%%%%%%%%%%%%%%%%%%%%%%%%%%%%%%%%%%%%%%%%%%%%%%%%%%%%%%%%%%%%%%%%%%%%%%%%%%%%%%%%%%%%%%%%%%%%%%%%%
\subsection{Recommendation performance over time}

To better understand the differences in performance between learned embedding (\texttt{SASRec}) and computed representations (\texttt{FELRec}), we plot the recommendation performance depending on the number of users who have interacted with the ground-truth item (Figure \ref{fig:item-rank}). First, we group user-item interactions by the number of interactions that the item has had with users so far. Next, we average the performance in every group. We plot the resulting performance curves separately for the new items partition (I) and the two remaining partitions (O+U). 

The results indicate that the more users interact with an item, the more accurately it is recommended. \texttt{FELRec} requires more interactions than \texttt{SASRec} to compute useful representations of items. On MovieLens 25M, \texttt{FELRec} performs similarly to \texttt{SASRec} after about $1000$ interactions. On Twitch, the performance of \texttt{SASRec} is substantially better all the time. We believe that \texttt{SASRec} performs better because it continuously improves item embeddings over the course of training, whereas \texttt{FELRec} must constantly compute representations from scratch as its cache gets cleared.

Interestingly, the recommendation performance of \texttt{FELRec} on the new items partition improves significantly over time compared to other partitions. While this phenomenon may seem counterintuitive given the comparatively lower performance reported by the metrics, most new items have little interactions, and the performance metrics reflect their larger contribution.

%%%%%%%%%%%%%%%%%%%%%%%%%%%%%%%%%%%%%%%%%%%%%%%%%%%%%%%%%%%%%%%%%%%%%%%%%%%%%%%%%%%%%%%%%%%%%%%%%%%%%%%%%
\subsection{Generalization to unseen data}

The dynamic cache allows \texttt{FELRec} to easily transfer knowledge between recommendation tasks. Since the cache stores arbitrary representations of users and items, \texttt{FELRec} can make recommendations for new, previously unseen, data without any further adjustments. Sequential recommenders, on the other hand, require either side information or finetuning to represent new items. This problem is exacerbated in the case of \textit{zero-shot learning} when a model runs inference on a dataset never seen before, and finetuning is not allowed. We treat zero-shot predictions as an extreme case of cold-start where no prior information exists.

Recall that during evaluation, we keep the existing representations in the cache and continue to compute new ones for test data. Therefore, by evaluating the performance of our model, we also evaluate the quality of representations that were computed during the training phase. In the first step towards zero-shot learning, we clear the cache completely and let \texttt{FELRec-Q} compute all representations from scratch. We refer to these results as \texttt{FELRec-Q-reset}. On MovieLens 25M (Table \ref{table-ml-25m}), \texttt{FELRec-Q} performs better than \texttt{FELRec-Q-reset} by $0.038$ rank and $14.76\%$ HR@10. On Twitch (Table \ref{table-twitch}), \texttt{FELRec-Q} is better by $0.028$ rank and $7.36\%$ HR@10. The results show that the quality of representations computed during training matters, but \texttt{FELRec} can compute new representations from scratch if necessary. 

We now evaluate the recommendation performance in zero-shot settings, i.e., when the model has not seen the data beforehand. Again, we clear the cache and let \texttt{FELRec-Q}, which was trained on MovieLens 25M (upstream dataset), compute representations of users and items from the Twitch (downstream) dataset, and vice versa. We refer to these results as \texttt{FELRec-Q-zero}. On Twitch (Table \ref{table-twitch}), \texttt{FELRec-Q-zero} performs better than \texttt{FELRec-Q-reset} by $2.67\%$ HR@10, but its rank is worse by $0.002$. On MovieLens 25M (Table \ref{table-ml-25m}), \texttt{FELRec-Q-zero} performs worse by $0.008$ rank and $2.84\%$ HR@10. In general, the better a model performs on the upstream dataset, the better its zero-shot performance on the downstream dataset. The results indicate that \texttt{FELRec} can compute representations from unknown data sources to solve the recommendation task. It leverages the underlying similarities between interaction datasets to generalize to unseen data. 

\begin{table}[t]
\caption{Recommendation performance in ablation studies on MovieLens 25M.}
\label{table-ablation}
\vskip 0.15in
\begin{center}
\begin{small}
\begin{sc}
\begin{tabular}{lcc}
\toprule
\multirow{2}{*}{model} & \multicolumn{2}{c}{total} \\
                       & rank       & hr@10        \\
\midrule
\texttt{FELRec-Q}                 & 0.052      & 87.85\%      \\
\midrule
\texttt{FELRec-Q-LN}              & 0.103      & 63.12\%      \\
\texttt{FELRec-P-LN}              & 0.118      & 54.54\%      \\
\texttt{FELRec-Q-noMLP}           & 0.163      & 12.34\%      \\
\texttt{FELRec-Q-shareMLP}        & 0.057      & 86.54\%      \\
\texttt{FELRec-Q-noType}          & 0.086      & 72.47\%      \\
\midrule
\texttt{FELRec-Q-NN}              & 0.052      & 87.90\%      \\
\bottomrule
\end{tabular}
\end{sc}
\end{small}
\end{center}
\vskip -0.1in
\end{table}

%%%%%%%%%%%%%%%%%%%%%%%%%%%%%%%%%%%%%%%%%%%%%%%%%%%%%%%%%%%%%%%%%%%%%%%%%%%%%%%%%%%%%%%%%%%%%%%%%%%%%%%%%
\subsection{Embedding \& complexity} 

Embedding layers constitute the majority of parameters in recommender systems. Sequential recommenders learn embeddings of items, factor models additionally learn embeddings of users. The complexity of these models is linearly correlated with the number of users and items in the dataset---the larger the dataset, the larger the model trained on it.

\texttt{FELRec} does not use any embedding layers, so its complexity does not depend on the dataset. The number of parameters remains constant because only the encoder and MLPs contribute to it. \texttt{FELRec} has about ten times fewer parameters than \texttt{SASRec} on MovieLens 25M (Table \ref{table-ml-25m}) and about $100$ times fewer parameters on Twitch (Table \ref{table-twitch}). This emphasizes the difference between \texttt{FELRec} and sequential recommenders with embedding layers in terms of what the model learns. Whereas \texttt{SASRec} learns a specialized embedding of a single dataset, \texttt{FELRec} learns a general embedding of interaction data, as demonstrated by its ability to generalize to new datasets.

\begin{figure}[t]%
    \centering
    \subfloat[\texttt{FELRec}]{{\includegraphics[width=0.9\columnwidth]{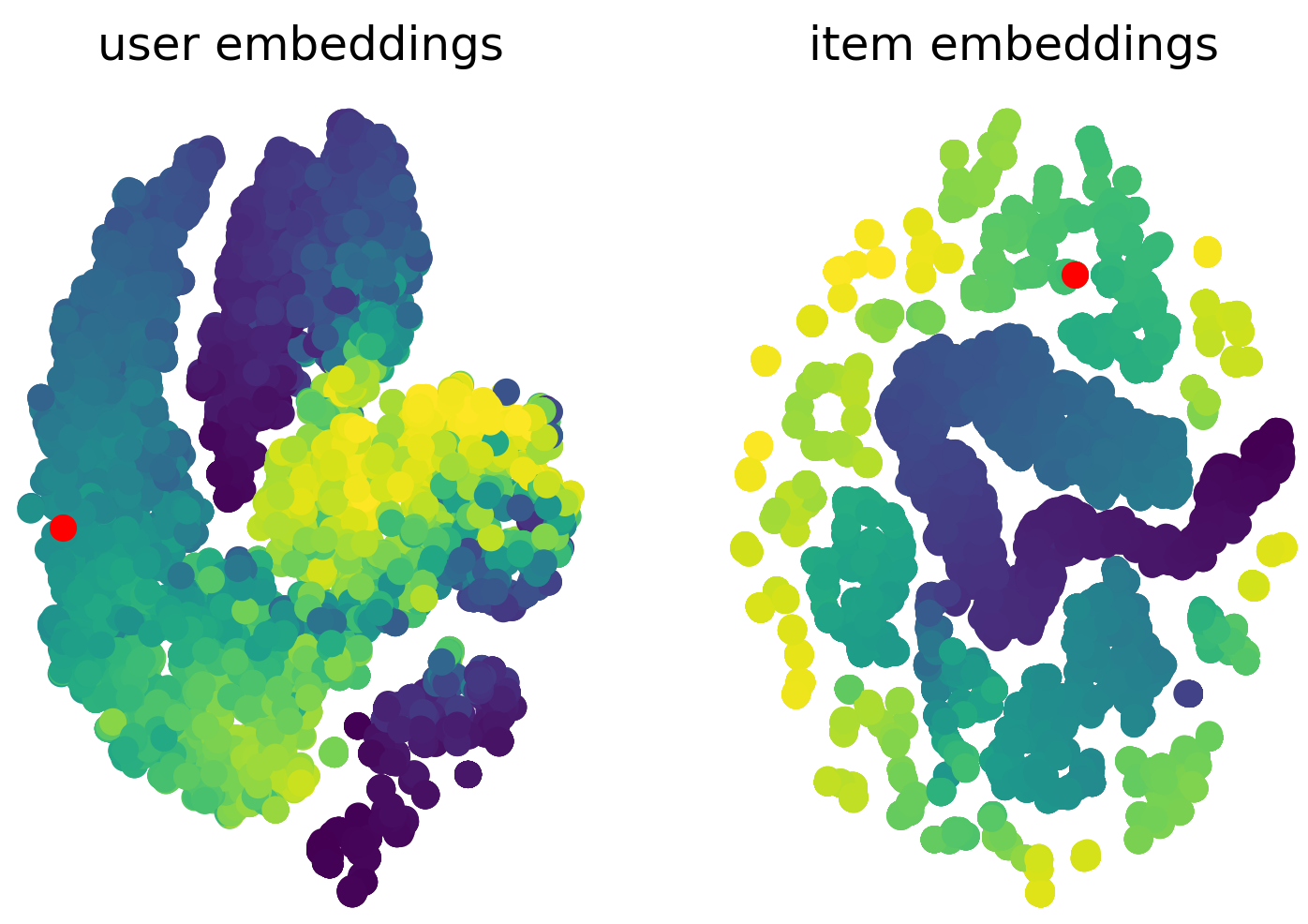} }}%
    \qquad
    \subfloat[\texttt{SASRec}]{{\includegraphics[width=0.9\columnwidth]{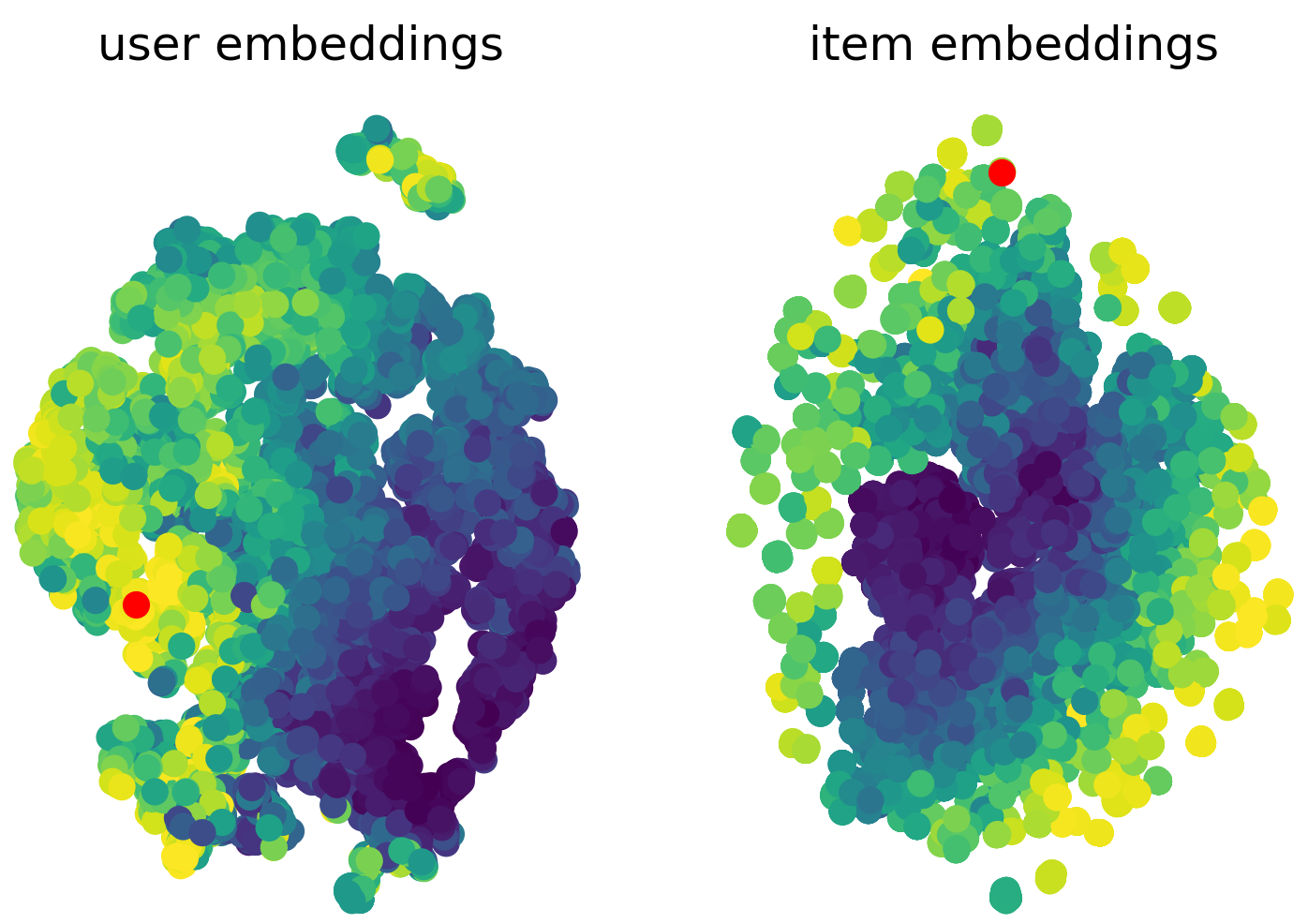} }}%
    \caption{T-SNE \cite{vandermaaten2008tsne} embedding of users and items represented by \texttt{FELRec} (a) and \texttt{SASRec} (b). Red dots are user (left) and item (right) in a single interaction. The embedding is coloured based on how well the users or items match the red dot on the opposing plot; bright colors mean a better match and dark colors a worse match. The ground-truth item (right dot) lies in the top $10\%$ recommended items.}%
    \label{fig:embedding}%
    \vskip -0.2in
\end{figure}

%%%%%%%%%%%%%%%%%%%%%%%%%%%%%%%%%%%%%%%%%%%%%%%%%%%%%%%%%%%%%%%%%%%%%%%%%%%%%%%%%%%%%%%%%%%%%%%%%%%%%%%%%
\subsection{Ablation studies}

We conduct several ablation studies on MovieLens 25M to evaluate the critical components of our proposed method and validate various architectural choices. These studies are designed to measure changes in recommendation performance when key elements are modified. The results are summarized in Table \ref{table-ablation}.

Our first study addresses the role of batch statistics in enhancing the performance of \texttt{FELRec}. Prompted by a prior report \cite{fetterman2020understanding} (addressed by \cite{richemond2020byol}) suggesting the crucial role of batch normalization for achieving optimal results, we examined its effectiveness compared to layer normalization, which does not rely on batch statistics. We implemented layer normalization in both the projection and prediction networks of \texttt{FELRec}, creating two variants: \texttt{FELRec-Q-LN} and \texttt{FELRec-P-LN}. The original model, \texttt{FELRec-Q}, outperformed these variants by $0.051$ and $0.066$ rank ($24.73\%$ and $33.31\%$ HR@10), respectively. These findings confirm that batch normalization significantly enhances performance compared to other normalization techniques.

The second ablation study investigates the impact of projection networks on recommendation performance. Given that these networks are often considered essential for achieving good performance \cite{chen2020improved}, we tested the performance of our model without them, to which we refer as \texttt{FELRec-Q-noMLP}. \texttt{FELRec-Q} outperforms this new variant by $0.111$ rank and $75.51\%$ HR@10. The results confirm that these MLPs are very important for the model to learn the task. In particular, this is highlighted by the HR@10 performance of \texttt{FELRec-Q-noMLP}, which is only marginally better than random recommendations. In a related experiment, we also tested parameter sharing within the projection networks (\texttt{FELRec-Q-shareMLP}). The results demonstrated that \texttt{FELRec-Q} only slightly outperformed \texttt{FELRec-Q-shareMLP}, with a rank improvement of $0.005$ and a HR@10 improvement of $1.31\%$. This result illustrates the expressive power of MLPs even when parameters are shared.

Finally, our third ablation study focuses on assessing the significance of including type embedding within our model architecture. This embedding is designed to help the model differentiate between types of input sequences, such as users versus items. We aimed to determine if this additional explicit information contributes positively to model performance, or if the model can adequately infer these differences without explicit type embeddings. To this end, we created a variant of the model, \texttt{FELRec-Q-noType}, by removing the type embedding. Comparisons show that \texttt{FELRec-Q} outperforms \texttt{FELRec-Q-noType} by $0.034$ in rank and $15.38\%$ HR@10. These results confirm that the inclusion of type embedding is crucial, providing valuable explicit information that enhances the overall performance of the model.

%%%%%%%%%%%%%%%%%%%%%%%%%%%%%%%%%%%%%%%%%%%%%%%%%%%%%%%%%%%%%%%%%%%%%%%%%%%%%%%%%%%%%%%%%%%%%%%%%%%%%%%%%
\subsection{Popularity bias}

Throughout the experiments, we also compared a number of 2D projections of learned embedding (\texttt{SASRec}) and computed representations (\texttt{FELRec}), hoping we would find the same clusters of users that share similar behavior. We discovered that both \texttt{FELRec} and \texttt{SASRec} suffer from \textit{popularity bias} \cite{cremonesi2010performance}, a tendency of a model to recommend the most popular items to the majority of users, often in disregard of their preferences.

To visualize the embeddings, we sample $10,000$ user-item interactions from MovieLens 25M, compute the representations of users and items, fit users and items separately into a 2D embedded space using t-SNE (t-distributed Stochastic Neighbor Embedding) using the standard parameters from the original paper \cite{vandermaaten2008tsne}, and plot a heat map that visualizes the similarity between embeddings of users and items (as predicted by the model). 

Figure \ref{fig:embedding} shows one such heat map produced for a single user-item interaction (red dots). The embedding is colored based on how well the users (left) or items (right) match the red dot on the opposing plot; bright colors mean a better match and dark colors a worse match. The ground-truth item (right dot) lies within the top $10\%$ of recommended items (i.e., HR@10\%).

In the embedding of both models, we identified several regions of items that are recommended to the majority of randomly chosen users, i.e., these regions are bright for most users. Likewise, some regions are almost never active (they roughly correspond to the dark-coloured regions in Figure \ref{fig:embedding}). A more thorough study of popularity bias could be of interest, but it is out of the scope of this work.

%%%%%%%%%%%%%%%%%%%%%%%%%%%%%%%%%%%%%%%%%%%%%%%%%%%%%%%%%%%%%%%%%%%%%%%%%%%%%%%%%%%%%%%%%%%%%%%%%%%%%%%%%
\subsection{Nearest neighbor recommendation}

We can find users that share similar interests by comparing their representations with the cosine similarity measure. Consequently, \texttt{FELRec} performs well if we make recommendations to the nearest neighbors of a user as her proxy. Specifically, we find ten nearest neighbors of a user with cosine similarity. Next, we recommend items to the neighbors and average the scores. The recommendation results, to which we refer as \texttt{FELRec-Q-NN} (Table \ref{table-ablation}), show that nearest neighbor recommendation can slightly improve the performance of our model.

%%%%%%%%%%%%%%%%%%%%%%%%%%%%%%%%%%%%%%%%%%%%%%%%%%%%%%%%%%%%%%%%%%%%%%%%%%%%%%%%%%%%%%%%%%%%%%%%%%%%%%%%%
\section{Discussion}
%%%%%%%%%%%%%%%%%%%%%%%%%%%%%%%%%%%%%%%%%%%%%%%%%%%%%%%%%%%%%%%%%%%%%%%%%%%%%%%%%%%%%%%%%%%%%%%%%%%%%%%%%
\label{sec:discussion}

Our experiments show that \texttt{FELRec} performs well in item cold-start scenarios with just user-item interactions and no side information. We demonstrate that our approach processes new interactions to improve existing representations of users and items without complex optimization techniques such as gradient descent---a single forward pass over a sequence of existing representations is enough. Further, our model can represent new users and new items in zero-shot settings based on interaction data from previously unseen sources.

However, in the current form, our model also has some limitations that open up future research possibilities. For example, other approaches might be better suited if side information is available. In this particular situation, side information can offer a rich training signal, so the models that use it might not require complex representation learning approaches such as the ones we use.

Another point worth mentioning is that \texttt{FELRec} requires considerably large datasets to learn the recommendation task well. During prototyping, our model struggled in the initial experiments when trained only on small interaction datasets. We found that training a good model requires datasets of at least $20$ million interactions, which exist in multiple application domains.

%%%%%%%%%%%%%%%%%%%%%%%%%%%%%%%%%%%%%%%%%%%%%%%%%%%%%%%%%%%%%%%%%%%%%%%%%%%%%%%%%%%%%%%%%%%%%%%%%%%%%%%%%
\section{Conclusion}
%%%%%%%%%%%%%%%%%%%%%%%%%%%%%%%%%%%%%%%%%%%%%%%%%%%%%%%%%%%%%%%%%%%%%%%%%%%%%%%%%%%%%%%%%%%%%%%%%%%%%%%%%
\label{sec:conclusion}

Recommender systems operate in dynamic environments where they must deal with the cold-start problem, a phase when a recommender lacks sufficient information to make a good recommendation to the user. In this paper, we have proposed a new paradigm for training sequential recommenders to address item cold-start. The main idea is to replace embedding layers with a dynamic storage that can keep an arbitrary number of representations of users and items and has no trainable weights. In contrast to similar approaches, our model represents new users and items without side information and the time-consuming process of finetuning, instead it runs a single forward pass over a sequence of existing representations. In item cold-start scenarios, our method outperforms similar method by $29.50\%$--$47.45$\%. Further, it generalizes well to previously unseen large datasets without additional training. We believe that our proposed method can be useful for the improvement of large online recommender systems.

%%%%%%%%%%%%%%%%%%%%%%%%%%%%%%%%%%%%%%%%%%%%%%%%%%%%%%%%%%%%%%%%%%%%%%%%%%%%%%%%%
%%%%%%%%%%%%%%%%%%%%%%%%%%%%%%%%%%%%%%%%%%%%%%%%%%%%%%%%%%%%%%%%%%%%%%%%%%%%%%%%%
%%%%%%%%%%%%%%%%%%%%%%%%%%%%%%%%%%%%%%%%%%%%%%%%%%%%%%%%%%%%%%%%%%%%%%%%%%%%%%%%%

\backmatter

\bmhead{Supplementary information}

The article contains no further supplementary files.

\bmhead{Acknowledgments}

This work was funded by the German Ministry for Education and Research (BMBF) within the Berlin Institute for the Foundations of Learning and Data - BIFOLD (project grants 01IS18025A and 01IS18037I) and the Forschungscampus MODAL (project grant 3FO18501).

\section*{Declarations}

\bmhead{Conflict of interest} The authors have no competing interests to declare that are relevant to the content of this article.

\bmhead{Code availability} The source code is publicly available at\\ \href{https://github.com/kweimann/FELRec}{https://github.com/kweimann/FELRec}.

\begin{appendices}

\section{Baseline methods}
\label{app:baselines}

We made changes to the original implementations of baseline methods to adapt them to our recommendation tasks. We briefly discuss these changes below.

\begin{itemize}
    \item \texttt{BPR-MF}. We train the model using stochastic gradient descent (SGD) with momentum $m=0.9$. We use weight decay of $0.001$.

    \item \texttt{GRU4Rec}. We follow the original implementation with some changes: $N=3$ GRU layers, BPR loss and no session-parallel mini-batching. Instead, we cache user sessions (hidden states) and use the cached session state whenever the corresponding session appears in a batch. This gives us more freedom in picking sessions for a batch. As a result, every batch during the training epoch is composed of a randomly chosen subset of unfinished sessions.
    
    \item \texttt{GRU4Rec-BPTT}. In comparison to \texttt{GRU4Rec}, we use an item embedding layer instead of 1-of-N encoding, and we have no activation in the prediction layer (as opposed to \textit{tanh}). Further, we add a linear layer after the last GRU layer and share the weights of the prediction layer with the item embedding. We compute the BPR loss at every position in the output sequence. We pad empty positions in the input sequence with zeros.
    
    \item \texttt{SASRec}. We follow the proposed model architecture with some changes: $N=3$ self-attention blocks, positional encoding instead of learned positional embedding and BPR as the loss function (instead of binary cross entropy). The architecture of the self-attention blocks is identical to the encoder in our \texttt{FELRec}. We add a linear layer before the prediction layer. The prediction layer shares parameters with the item embedding layer.
    
    \item \texttt{JODIE}. We noticed that the t-Batch algorithm substantially increases training times and memory consumption on large datasets. We replace t-Batch with the same naive batching strategy that we use in our model. To further reduce memory consumption, we directly feed the learnable static embeddings to the model instead of one-hot encoded vectors. We also dropped the feature embeddings because we do not use side information. 
\end{itemize}

All baseline models share the same embedding and latent dimension size of $128$. \texttt{BPR-MF}, \texttt{GRU4Rec}, \texttt{GRU4Rec-BPTT} and \texttt{SASRec} are trained using stochastic gradient descent (SGD) with momentum $m=0.9$ and learning rate between $0.1$ and $0.001$ for $50$ epochs ($100$ in case of \texttt{BPR-MF}). \texttt{GRU4Rec-BPTT} and \texttt{SASRec} take up to $64$ items in the input sequence. We use between $1$ and $32$ negative samples for BPR. Dropout is set to $0.1$. Batch size is $1024$ ($4096$ for \texttt{BPR-MF}). 

\end{appendices}


\bibliographystyle{IEEEtran}
\bibliography{sn-bibliography}% common bib file

\end{document}